\documentclass[twocolumn,tighten,times,twocolappendix]{aastex631}

\usepackage{amssymb,amsmath}
\usepackage{bm}
\usepackage{comment}

\shortauthors{Schad et al.}

\newcommand{\ex}{\hat{e}_x}
\newcommand{\ey}{\hat{e}_y}
\newcommand{\ez}{\hat{e}_z}
\newcommand{\expr}{\hat{e}_x^\prime}
\newcommand{\eypr}{\hat{e}_y^\prime}
\newcommand{\ezpr}{\hat{e}_z^\prime}
\newcommand{\pp}{\phantom{-}}
\newcommand{\uat}[2]{\href{http://astrothesaurus.org/uat/#2}{#1 (#2)}}

\received{2022 April 17}
\revised{2022 May 17}
\accepted{2022 May 17}

\graphicspath{{./}}
\defcitealias{landidegl2004}{LL04}
\defcitealias{schad2021}{Paper I}

\begin{document}
\title{\Large{Thomson scattering above solar active regions and an \\ ad-hoc polarization correction method for the emissive corona}}

\correspondingauthor{Thomas A. Schad}
\email{tschad@nso.edu}
\author[0000-0002-7451-9804]{Thomas A. Schad}
\affiliation{National Solar Observatory, 22 Ohia Ku Street, Pukalani, HI 96768, USA}
\author[0000-0001-5459-2628]{Sarah A. Jaeggli}
\affiliation{National Solar Observatory, 22 Ohia Ku Street, Pukalani, HI 96768, USA}
\author[0000-0002-6003-4646]{Gabriel I. Dima}
\affil{Cooperative Institute for Research in Environmental Sciences, CU Boulder, CO 80309, USA}
\affil{NOAA National Centers for Environmental Information, DSRC, 325 Broadway, Boulder, CO 80305, USA}

\begin{abstract}
Thomson scattered photospheric light is the dominant constituent of the lower solar corona's spectral continuum viewed off-limb at optical wavelengths.  Known as the K-corona, it is also linearly polarized.  We investigate the possibility of using the \textit{a priori} polarized characteristics of the K-corona, together with polarized emission lines, to measure and correct instrument induced polarized crosstalk.  First we derive the Stokes parameters of Thomson scattering of unpolarized light in an irreducible spherical tensor formalism.  This allows forward synthesis of the Thomson scattered signal for the more complex scenario of symmetry-breaking features in the incident radiation field, which could limit the accuracy of our proposed technique.  For this, we make use of an advanced 3D radiative magnetohydrodynamic coronal model.  Together with synthesized polarized signals in the \ion{Fe}{13} 10746 \mbox{\AA} emission line, we find that an ad hoc correction of telescope and instrument induced polarization crosstalk is possible under the assumption of a non-depolarizing optical system. 
\end{abstract}

\section{Introduction} \label{sec:intro}

The accurate measurement of polarized coronal emission lines and continua is predicate to their use as remote diagnostics of the solar coronal magnetic field \citep[see, \textit{e.g.}][]{judge2013,landi2016}.  Frontier large-aperture coronagraphic facilities, like the 4 m National Science Foundation's Daniel K. Inouye Solar Telescope \citep[DKIST:][]{rimmele2020} and the 1.5 m Coronal Solar Magnetism Observatory \citep[COSMO:][]{tomczyk2016}, are designed to conduct highly sensitive off-limb measurements of the scattering- and Zeeman-effect-induced polarized signals at visible and infrared wavelengths, where the intensity contrast relative to the solar disk is ${<}10^{-5}$.  Achieving accurate polarimetric calibration of such large-aperture systems presents numerous challenges \citep[see, \textit{e.g.},][]{harrington2017JATIS}, especially in the absence of calibration optics that can extend across the entire entrance pupil.  This motivates the development of alternative methods that seek to calibrate or validate system performance through the measurement of natural sources that are otherwise well characterized (\textit{e.g.}, standard stars) or whose formation is sufficiently understood.  Examples of the latter include daytime sky Rayleigh scattering \citep{harrington2017} and polarized Zeeman profiles of solar surface magnetic fields \citep{sanchez1992, kuhn1994, schlichenmaier2002}. Here we introduce a technique for ad-hoc polarization correction (or validation) that utilizes the expected polarized characteristics of the off-limb corona, in particular, its near Sun continuum (K-) and line emissive (E-) constituents.  

Within the lower corona (\textit{i.e.}, heliocentric distances, or elongations, $\lesssim$2$R_{\odot}$ as viewed from Earth), the off-limb continuum is dominated by Thomson scattering of photospheric light by free coronal electrons \citep[\textit{i.e.}, the K-corona,][]{schuster1879, minnaert1930, vanDeHulst1950, inhester2015}.  It has radially decreasing intensities, peaking at a few millionths of the disk intensity, and it is linearly polarized with amplitudes of 10 to 70\%, increasing radially.  Eclipse observations by \citet{vorobiev2020} show the K-corona polarization on large spatial scales is oriented tangential to the solar limb within measurement uncertainties.  This is as expected for Thomson scattering of an unpolarized photospheric radiation field that is cylindrically-symmetric relative to the radial direction, whereas the presence of symmetry breaking features like sunspots can induce deviations in the polarization direction,  as discussed in this work and by \citet{Saint_Hilaire_2021}.  In addition, the K-corona is largely spectral line free as the high thermal velocity of coronal electrons smooths out all but the strongest (Fraunhofer) absorption lines \citep{cram1976}.  

In contrast, the dust-scattered F-corona (\textit{i.e.} the inner zodiacal light) does preserve spectral features of the incident radiation.  Utilizing this to help separate the F and K corona, in a manner similar to \citet{vanDeHulst1950}, the semi-empirical eclipse model of \citet{blackwell1966} finds the inner F-corona polarization to be very weak, decreasing from 0.9\% at elongations of 20 R$_{\odot}$ to 0.05\% at 5 R$_{\odot}$ \citep[see also][]{ingham1961}.  This lends support for assuming the F-corona has negligible polarization near the Sun, which subsequently allows F- and K-corona separation through polarized measurements \citep{koutchmy1985}.  Using color-dependent analysis, however, \citet{boe2021} recently found larger fractions of F-corona scattering at small elongations when compared to \citet{koutchmy1985}, which may result from the relative uncertainty in our knowledge of the F-coronal polarization, as discussed by \citet{mann1992}, \citet{kimura1998}, and \citet{lamy2021}, and highlights the importance of spectropolarimetric measurements.  

While recognizing the above uncertainties, it remains the case that the off-limb continuum polarization is substantially polarized and oriented, with few exceptions, tangential to the projected radial vector.  As a result, these characteristics offer \textit{a priori} constraints on the polarized response of a coronagraph, provided some additional knowledge of the optical system.  \citet{lamy2021}, for example, minimized the deviation angle of the polarization orientation from the tangential direction by adjusting the polarized transmissions of a single linear polarizer model for LASCO-C3 observations. 

We consider the case of a more generalized optical system with a larger number of free variables, as required especially to model articulated full-Stokes polarimeters like DKIST.  \citet{jaeggli2022} have argued for the benefits of modeling optical systems using the polar decomposition for a non-depolarizing Mueller matrix, which treats an arbitrary optical system as a non-depolarizing combination of an elliptical diattenuator and an elliptical retarder, as described in \citet{chipman2018}. As such it can treat crosstalk from intensity to polarized states as well as between the polarized states.  As will be shown, applying this model to coronal observations requires additional constraints.  In addition to the continuum polarization orientation, we take advantage of the spectral line-free character (apart from telluric absorption) of the Thomson scattered continuum and the presumption that the F-corona is very weakly polarized in the inner corona.  An additional constraint is the expected zero net circular polarization (Stokes V) within forbidden emission-lines.  Under conditions of excitation by unpolarized photospheric radiation and/or thermal collisions, the circularly polarized profile results from Zeeman splitting and is integrated along a line-of-sight traversing the diffuse, optically-thin corona (see, \textit{e.g.}, \citet{schad2020}).  In absence of other mechanisms, the line-integrated Stokes V emissivity is zero. 

Below, we demonstrate the use of the K and E coronal signals to constrain the polarimetric system response using a non-depolarizing optical model applied to synthetically calculated observables.  As the upcoming large aperture facilities, \textit{esp.} DKIST, will often target active regions with limited field-of-views, we consider possible deviations in the tangential character of Thomson polarization at low heights near symmetry-breaking features.  The role of symmetry-breaking in the incident radiation field on polarized emission lines has already been treated in our prior work (\citet{schad2021}, hereafter \citetalias{schad2021}).  We find that jointly considering the K- and E- coronal signals can benefit from a common formalism for specifying the incident radiation field, complementary to the methods introduced by \citet{Saint_Hilaire_2021}.  Therefore, we begin in Section~\ref{sec:thomson} by deriving the polarized emissivities in the Stokes formalism for Thomson scattering while making use of the irreducible spherical tensors introduced in \citet{landidegl1983}.  In Section~\ref{sec:forward}, we forward synthesize Stokes spectra through a 3D radiative magnetohydrodynamic coronal model of a bipolar active region before discussing and demonstrating the ad hoc polarization correction technique in Section~\ref{sec:adhoc}. 

\section[Thomson Scattering of Unpolarized Light]{Thomson Scattering of Unpolarized Light}\label{sec:thomson}

\begin{figure}
    \centering
    \includegraphics[width=0.95\columnwidth]{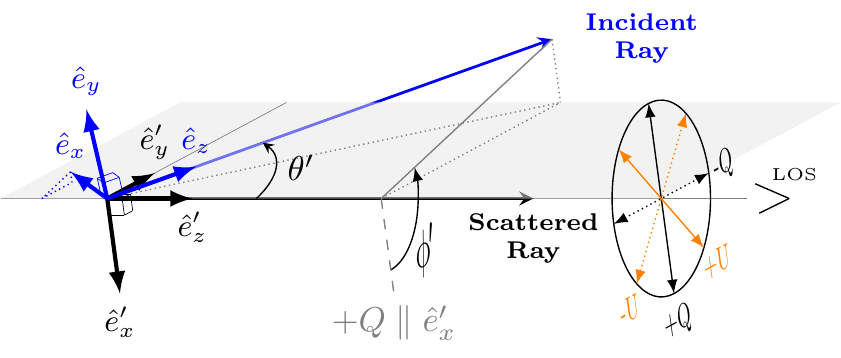} 
    \caption{Geometry of the Thomson scattering calculation referenced to a fixed Stokes frame along the line-of-sight (LOS) defined by the scattering direction and with Stokes +Q aligned parallel to the $\hat{e}^{\prime}_{x}$ polarization vector of the scattered ray.}
    \label{fig:thomson_geometry}
\end{figure}

The classical differential cross section for Thomson scattered radiation resultant from the interaction of an incident plane wave with a single electron is given by \citep[see, \textit{e.g.},][]{jackson1975}
\begin{equation}
\left. \frac{d \sigma}{d \Omega^{\prime}} \right |_{\bm{k}^{\prime}} = r_{e}^{2} \left| \bm{\epsilon^{\prime*}} \cdot \bm{\epsilon} \right|^{2}  \label{eqn:cross_section}
\end{equation}
where $\bm{\epsilon}^{\prime}$ and $\bm{\epsilon}$ denote the scattered and incident polarization vectors for waves propagating along $\bm{k}^\prime$ and $\bm{k}$, respectively. $r_{e}$ is the classical electron radius ($2.82 \times 10^{-15}$ m), and the asterisk denotes complex conjugation.  The differential cross section is the power radiated per unit solid angle (optionally per unit wavelength) into the scattering direction per unit incident flux, \textit{i.e.} power per unit area and optionally per unit wavelength.  The prime notation on $d \Omega^{\prime}$ is used to clarify that the unit solid angle here is specified in the scattering frame directed along $\bm{k}^\prime$.

\subsection{Polarized Emissivities in a Fixed Stokes Frame}\label{sec:fixed_frame} 

Calculating the polarized emission coefficients (\textit{i.e.}, the spectral radiance per unit volume) of Thomson scattering along a given line of sight requires considering a volume element with electron density $n_{e}$ and integrating over all incident radiation field directions.  It is beneficial to consider a fixed geometry along the scattering direction with a fixed orientation of the Stokes vectors along the observer's line-of-sight.  In Figure~\ref{fig:thomson_geometry}, two Cartesian coordinate frames are defined with a common origin at the scattering location.  The scattered ray lies along the z axis in the primed coordinate frame (\textit{i.e.}, parallel to the unit vector $\hat{e}^{\prime}_{z}$).  Its polarization unit vectors, which are orthogonal to the scattering direction, are given by $\hat{e}^{\prime}_{x}$ and $\hat{e}^{\prime}_{y}$.  Meanwhile, the incident ray's direction is free to vary and has a polar angle $\theta^\prime$ and an azimuthal angle $\phi^{\prime}$ in the coordinate frame fixed to the scattering direction.  The second Cartesian frame is aligned with the incident ray, with $\hat{e}_{z}$ being parallel to its direction and $\hat{e}_{x}$ and $\hat{e}_{y}$ being its polarization unit vectors.  The two coordinate frames are related through a general three dimensional rotation.  For the case of unpolarized incident radiation, this can be further simplified as the rotation of the incident ray's polarization unit vectors about its propagation direction is unimportant.  Thus, this degree of freedom can be removed without losing generality.  We here specify the incident plane defined by $\hat{e}_{x}$ and $\hat{e}_{z}$ to be parallel to the scattered direction $\hat{e}^{\prime}_{z}$.  Subsequently, the sets of unit vectors are related according to $[\ex,\ey,\ez]^{T}  = \mathbb{R}_{y}(\theta^{\prime})\mathbb{R}_{z}(\phi^{\prime}) [\expr,\eypr,\ezpr]^{T} $
where $\mathbb{R}_{y}(\theta^{\prime})$ and $\mathbb{R}_{z}(\phi^{\prime})$ are \textit{passive} Cartesian transformation matrices for counter-clockwise axis rotation about the intrinsic y and z axes, \textit{i.e.}, \begin{equation}
    \begin{bmatrix}
    \ex \\
    \ey \\
    \ez \\
    \end{bmatrix} =
    \begin{pmatrix}
    C_{\theta^\prime} & 0 & -S_{\theta^{\prime}} \\
    0 & 1 & 0 \\
    S_{\theta^{\prime}}& 0 & C_{\theta^\prime} \\
    \end{pmatrix} 
    \begin{pmatrix}
    C_{\phi^\prime} & S_{\phi^\prime} & 0 \\
    -S_{\phi^\prime} & C_{\phi^\prime} & 0 \\
    0 & 0 & 1 \\
    \end{pmatrix}
    \begin{bmatrix}
    \expr \\
    \eypr \\
    \ezpr \\
    \end{bmatrix}
    \label{eqn:vector_transform_long}
\end{equation}
using the short-hand notation $C_{x} = \cos x$ and $S_{x} = \sin x$.

Assuming the incident radiation field is unpolarized, an individual wave can be decomposed into two equal components of linearly polarized light aligned with the $\ex$ and $\ey$ axes.  The unit incident flux in the direction $\vec\Omega$ is given by $I_{\vec\Omega}(d\Omega)$, or equivalently, $I_{\vec\Omega^\prime}(d\Omega^\prime)$ when specified in the geometry of the scattering frame for which $\vec\Omega^\prime$ is oriented at $(\theta^{\prime},\phi^{\prime})$.  Similar to \citet{kosowsky1996}, we can define the Stokes parameters of the scattered wave propagating in direction $\bm{k}^\prime$ ($\parallel \hat{e}^{\prime}_{z}$) using the canonical Stokes basis vectors in the primed coordinate frame and the cross section given in Equation~\ref{eqn:cross_section}.  The positive Stokes Q direction is chosen here to align with the $\hat{e}^{\prime}_{x}$ vector as shown in Figure~\ref{fig:thomson_geometry}.  After multiplying by the electron density $n_{e}$, the differential Stokes emissivities (\textit{i.e.} spectral radiance in the scattered direction $\hat{e}^{\prime}_{z}$ per unit volume per unit solid angle of the incident radiation) are given by
\begin{align}
\left. \frac{d\epsilon_{I}}{d \Omega^\prime} \right |_{\hat{e}^{\prime}_{z}} &= \frac{n_{e} r_{e}^2 I_{\vec\Omega^\prime}}{2} 
\left [ 
\sum_{i=x,y} \left | \expr \cdot \hat{e}_i \right | ^{2}
+ \sum_{i=x,y}  \left | \eypr \cdot \hat{e}_i \right | ^{2}
\right ]  \label{eqn:stokesI} \\ 
\left. \frac{d\epsilon_{Q}}{d \Omega^\prime} \right |_{\hat{e}^{\prime}_{z}} &= \frac{n_{e} r_{e}^2 I_{\vec\Omega^\prime}}{2} 
\left [ 
\sum_{i=x,y} \left | \expr \cdot \hat{e}_i \right | ^{2}
- \sum_{i=x,y}  \left | \eypr \cdot \hat{e}_i \right | ^{2}
\right ] \label{eqn:stokesQ} \\
\left. \frac{d\epsilon_{U}}{d \Omega^\prime} \right |_{\hat{e}^{\prime}_{z}} &= \frac{n_{e} r_{e}^2 I_{\vec\Omega^\prime}}{2} 
\left [ 
\sum_{i=x,y} \left | \hat{e}^{\prime}_a \cdot \hat{e}_i \right | ^{2}
- \sum_{i=x,y}  \left | \hat{e}^{\prime}_b \cdot \hat{e}_i \right | ^{2}
\right ]  \label{eqn:stokesU}
\end{align}
where $\hat{e}^{\prime}_a = (\expr + \eypr)/\sqrt{2}$ and $\hat{e}^{\prime}_b = (\expr - \eypr)/\sqrt{2}$ are the Stokes U basis vectors rotated $45^\circ$ relative to Stokes Q.  $\epsilon_{V}=0$ in the case of unpolarized incident radiation and is not considered further. Note that $d\Omega^\prime$ in Equations~\ref{eqn:stokesI}-\ref{eqn:stokesU} is the unit solid angle of the incident radiation in the scattering frame but not specifically in the scattering direction.  The unit solid angle of the scattered radiation is subsumed into the emission coefficient.

Using Equation~\ref{eqn:vector_transform_long}, we can transform the unit vectors onto a common basis to carry out the dot products in Equations~\ref{eqn:stokesI}-\ref{eqn:stokesU} which results in the Stokes emissivities of Thomson scattering along the scattered direction $\hat{e}^{\prime}_{z}$:
\begin{eqnarray}
\epsilon_{I}|_{\hat{e}^{\prime}_{z}} &=& \frac{n_{e} r_{e}^2}{2} \int d \Omega^{\prime}  I_{\theta^{\prime}\phi^{\prime}}^{\prime} \left( 1 + \cos ^{2} \theta^{\prime} \right ) \label{eqn:epsI_std} \\
\epsilon_{Q}|_{\hat{e}^{\prime}_{z}} &=& \frac{-n_{e} r_{e}^2}{2} \int d \Omega^{\prime} I_{\theta^{\prime}\phi^{\prime}}^{\prime} \sin ^{2} \theta^{\prime} \cos 2 \phi^{\prime}\label{eqn:epsQ_std} \\ 
\epsilon_{U}|_{\hat{e}^{\prime}_{z}} &=& \frac{-n_{e} r_{e}^2}{2}  \int d \Omega^{\prime} I_{\theta^{\prime}\phi^{\prime}}^{\prime} \sin ^{2} \theta^{\prime} \sin 2 \phi^{\prime} \label{eqn:epsU_std} 
\end{eqnarray}
where $I_{\vec\Omega^\prime}$ has been rewritten as $I_{\theta^{\prime}\phi^{\prime}}^{\prime}$ and is implicitly wavelength dependent. Note the sign differences in our equations for $\epsilon_{Q}$ and $\epsilon_{U}$ when compared with \citet{kosowsky1996} result from our definition of the $+Q$ reference direction and our use of $+\phi^\prime$ to represent the customary counter-clockwise azimuthal angle.  It is clear given the above that when $\phi^\prime=0$, $\epsilon_{Q}$ is negative and $\epsilon_{U}=0$, which corresponds to linear polarization perpendicular to the scattering plane.  

%%%%%%%%%%%%%%%%%%%%%%%%%%%%%%%%%%%%%%%%%%%%%%%%%%

\subsection[Quantifying the Incident Radiation Field with Irreducible Spherical Tensors]{Quantifying the Incident Radiation Field \\ with Irreducible Spherical Tensors}

While the above equations are complete, the radiation field is not conveniently quantified as its angular variation is referenced to the scattering direction.  The radiation field of an outer stellar atmosphere is more readily determined in a reference frame aligned with the radial direction, where in the nominal case, it is cylindrically-symmetric, non-polarized, and exhibits limb-darkening.  Similar to the case of the cosmic microwave background (CMB) described by \citet{kosowsky1996}, it is useful to expand the radiation field in an irreducible representation.  In the CMB field, spherical harmonics are often used.  Here, we employ the (KQ) representation introduced by \citet{landidegl1983} for which the components of the irreducible tensor of an unpolarized radiation field $J_{Q}^{K}$ at frequency $\nu$ are given by \citep[see Equation 5.157 of][hereafter LL04]{landidegl2004}:
\begin{eqnarray}
J_{0}^{0}(\nu) &=& \oint \frac{\mathrm{d} \Omega}{4 \pi} I(\nu, \vec{\Omega}) \label{eqn:J00}\\
J_{0}^{2}(\nu) &=& \frac{1}{2 \sqrt{2}} \oint \frac{\mathrm{d} \Omega}{4 \pi}\left(3 \cos ^{2} \theta-1\right) I(\nu, \vec{\Omega}) \label{eqn:J02} \\
J_{\pm 1}^{2}(\nu) &=& \mp \frac{\sqrt{3}}{2} \oint \frac{\mathrm{d} \Omega}{4 \pi} \sin \theta \cos \theta\ \mathrm{e}^{\pm \mathrm{i} \phi} I(\nu, \vec{\Omega}) \label{eqn:J12}\\
J_{\pm 2}^{2}(\nu) &=& \frac{\sqrt{3}}{4} \oint \frac{\mathrm{d} \Omega}{4 \pi} \sin ^{2} \theta\ \mathrm{e}^{\pm 2 \mathrm{i} \phi} I(\nu, \vec{\Omega}). \label{eqn:J22}
\end{eqnarray}
Note that our notation uses $\phi$ for the azimuthal angle instead of $\chi$ as used by \citetalias{landidegl2004}. These equations fully quantify the unpolarized radiation field within a given reference frame (\textit{i.e.}, in either the local stellar frame or within the scattering direction's frame), and they have the advantage that under rotations, the tensor transforms according to
\begin{equation}
J_{\phantom{\prime}Q}^{\prime K}
= \sum_{P} J_{P}^{K} \mathcal{D}_{P Q}^{K}(R)
\label{eqn:rotation}, 
\end{equation}
where $\mathcal{D}_{P Q}^{K}(R)$ is a rotation matrix discussed further below (see also ~\citetalias{landidegl2004} Equation 2.68 and Section 2.7.)

First, using Equations~\ref{eqn:J00}-\ref{eqn:J22}, the I, Q, and U emissivities in Equations~\ref{eqn:epsI_std}-~\ref{eqn:epsU_std} can be recast, after some algebra, as
\begin{align}    
\epsilon_{I}|_{\hat{e}^{\prime}_{z}} &= \frac{8\pi n_{e} r_{e}^2}{3} \left ( J_{0}^{\prime0} + \frac{J_{0}^{\prime2}}{\sqrt{2}} \right ) \label{eqn:epsI_tensor1} \\
             &= \sigma_{T} n_{e} J_{0}^{\prime0}  \left ( 1 + \frac{\omega^\prime}{2} \right ) \label{eqn:epsI_tensor2}\\
(\epsilon_{Q} + i \epsilon_{U})|_{\hat{e}^{\prime}_{z}} &= -\sqrt{3} \sigma_{T} n_{e} J_{2}^{\prime 2} \label{eqn:epsQU_tensor}
\end{align}
where the primed $J_{Q}^{\prime K}$ notation denotes spherical tensor components in the coordinate frame aligned with the scattering direction (\textit{i.e.}, along an observer's line-of-sight). The mean intensity ($K,Q=0,0$) is preserved in all coordinate frames ($J_{0}^{\prime0}  =  J_{0}^{0}$). The term $\omega$ (or $\omega^{\prime}$ in the scattered frame) is referred to as the anisotropy factor given by  $\sqrt{2}J_{0}^{2}/J_{0}^{0}$ (or $\sqrt{2}J_{0}^{\prime2}/J_{0}^{\prime0}$).  $\sigma_{T}$ ($= \frac{8\pi}{3}r_{e}^2 \approx 0.665 \times 10^{-24}$ cm$^{2}$) is the total Thomson electron cross section.  Equations~\ref{eqn:epsI_tensor2} and~\ref{eqn:epsQU_tensor} are comparable to the spherical-harmonic expansion used in the CMB theory \citep{kosowsky1996}. In particular, both sets of equations agree that linear polarization in the scattered light is directly proportional to the quadrupole moment ($K,Q = 2,2$) of the incident radiation field in the frame aligned with the scattered direction. 

\subsection{The Cylindrically-Symmetric Limb-Darkened Case}\label{sec:cyn_sym}

\begin{figure}
    \centering
    \includegraphics[width=0.95\columnwidth]{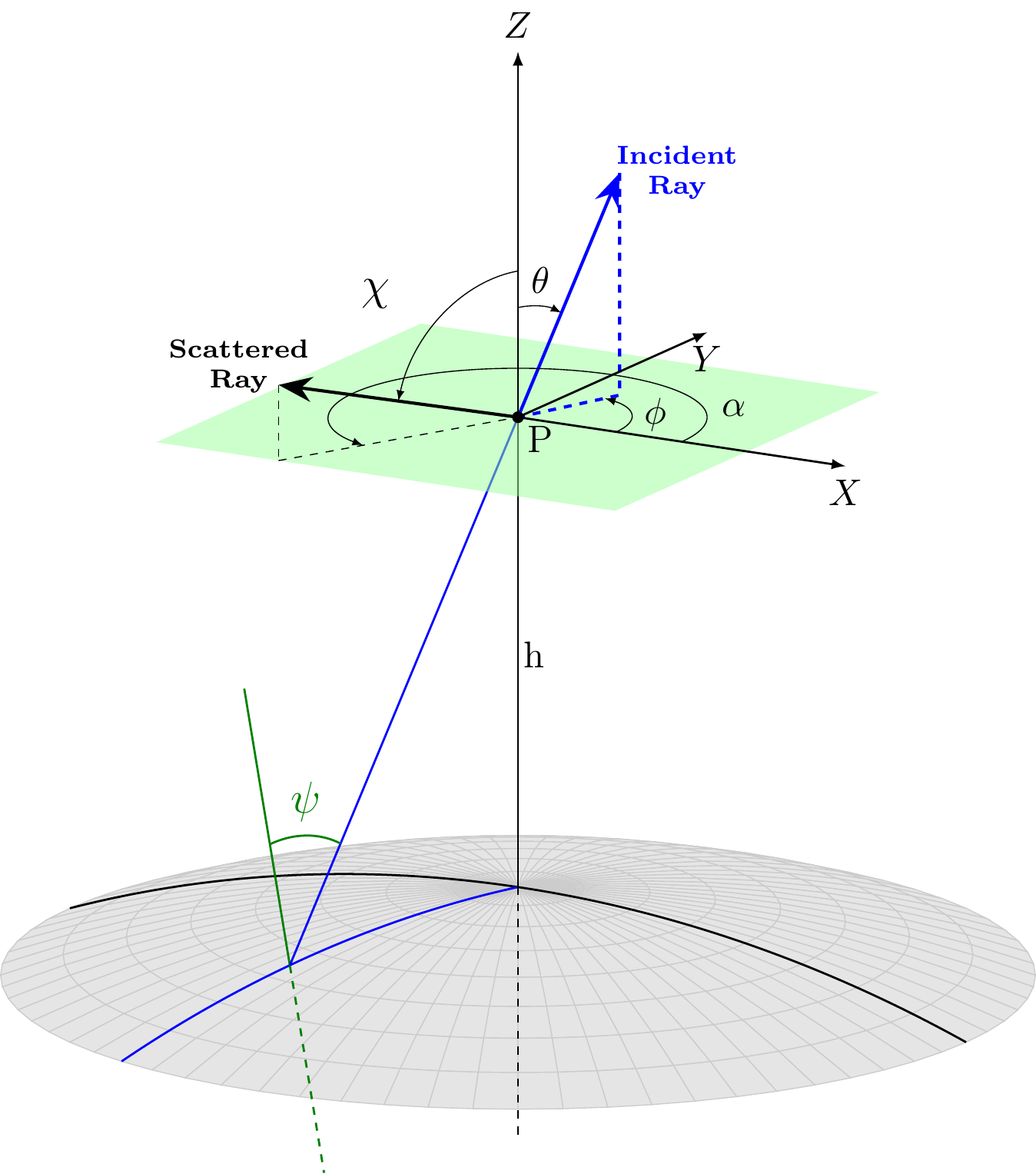}
    \caption{Geometry for the line-of-sight (LOS) frame relative to the axes used for quantifying the radiation field in the local radial frame.  The LOS frame is specified by the zyz extrinsic Euler angles $\gamma = 0$ (not shown), $\beta = \chi$ and $\alpha = \alpha$.  The choice $\gamma=0$ results in the +Q Stokes direction being aligned with the projected radial direction at point P, which is located at height h above the stellar surface.  One example incident ray is shown in blue, which exits the stellar photosphere at an angle $\psi$ relative to the surface normal.}
    \label{fig:radiation_field}
\end{figure}

Equations~\ref{eqn:epsI_tensor2} and~\ref{eqn:epsQU_tensor} can be used to compute the Thomson scattered emissivities for any unpolarized incident radiation field.  We first consider the typical stellar case of a unpolarized radiation field that is cylindrically symmetric relative to the star's radial direction, that varies in intensity relative to the polar angle due to limb-darkening, and is emergent into an optically thin outer atmosphere.  In such case, the incident radiation field at any point in the outer atmosphere is quantified entirely by two components ($J_{0}^{0}$ and $J_{0}^{2}$), but only in the radial coordinate frame where the +z-axis is directed radially outwards.  As this does not align with the scattered direction, we must transform these components, using Equation~\ref{eqn:rotation}, into the coordinate frame defined above for Equations~\ref{eqn:epsI_tensor2} and~\ref{eqn:epsQU_tensor}, \textit{i.e.} the line-of-sight is along its respective z-axis. 

$\mathcal{D}_{P Q}^{K}(R)$ in Equation~\ref{eqn:rotation} describes a Eulerian rotation with $R$ representing the ($\alpha,\beta,\gamma$) angle triad.  The order of active rotations are $\gamma$ about the original (``fixed") z-axis, $\beta$ about the original y-axis, and $\alpha$ about the original z-axis.  Transforming $J_{Q}^{K}$ from the radial frame to the primed coordinate axes that align with the line-of-sight frame requires a \textit{passive} rotation of the axes.  This implies a transformation consisting of transposed elemental rotation matrices\footnote{This is equivalent to active counterclockwise rotations in a right-handed coordinate system with the transformation ($\alpha \rightarrow -\alpha$, $\beta \rightarrow -\beta$, $\gamma \rightarrow -\gamma)$}.  In the cylindrically-symmetric case, the primary angle that modifies the amplitude of the polarized scattering is the inclination angle of the line-of-sight relative to the radial z-axis, which we denote $\chi$ and corresponds to $\beta$ in the axes rotation.  $\alpha$ and $\gamma$, in this case, control only the orientation of the Stokes reference frame (see Figure~\ref{fig:radiation_field}).  The passive rotation $R = (-\pi/2,\chi,0)$ transforms coordinate axes such that the z-axis is aligned along a line-of-sight directed in the -Y direction of the radial frame, and the x-axis (defining +Q) is aligned parallel with radial direction.  Using Equation~\ref{eqn:rotation} and the algebraic formula for the associated reduced rotation matrices, the radiation field tensor components in the line-of-sight frame (again, for the cylindrically symmetric case, where $J_{\pm 1}^{2} =0$ and $J_{\pm 2}^{2}=0$) are given by:
\begin{align}
J_{0}^{\prime 2} &= \frac{1}{2}(3\cos^2\chi-1) J_{0}^{2} \\
J_{2}^{\prime 2} &= \sqrt{\frac{3}{8}} \left( \sin^{2} \chi \right) J_{0}^{2}  = \frac{\sqrt{3}}{4} \omega J_{0}^{0} \sin^{2} \chi
\end{align}
and the anisotropy factor in the rotated frame is related to the radial frame by
\begin{equation}
    \omega^{\prime} = \frac{1}{2}(3\cos^2\chi-1) \omega. 
\end{equation}
Therefore, Equations~\ref{eqn:epsI_tensor2} and~\ref{eqn:epsQU_tensor} become 
\begin{align}    
\epsilon_{I}|_{\hat{e}^{\prime}_{z}} &= \sigma_{T} n_{e} J_{0}^{0}  \left ( 1 + \frac{\omega}{2} - \frac{3}{4}\omega \sin^{2} \chi \right )  \label{eqn:epsI_tensor_sym} \\
(\epsilon_{Q} + i \epsilon_{U})|_{\hat{e}^{\prime}_{z}} &= -\frac{3}{4}\sigma_{T} n_{e} \omega J_{0}^{0} \left ( \sin^{2} \chi \right ). \label{eqn:epsQU_tensor_sym}
\end{align}
As $J_{0}^{0}$ and $\omega$ in an outer atmosphere are both positive and real, $\epsilon_{U}=0$ and $\epsilon_{Q}$ is negative, which implies the linear polarization orientation is perpendicular to the projected radial direction, as expected.  

\subsubsection{Analytic Formulae for the Radiation Field Tensor}

Analytic formulae can be derived for $J_{0}^{0}$ and $\omega$ when expanding the limb-darkened stellar intensity in the canonical manner, as is thoroughly described in Section 12.3 of \citetalias{landidegl2004}.  We summarize this formalism here for completeness so to increase the utility of the above equations.  The limb-darkening law is first written as
\begin{equation}
    I(\psi) = I(0) \left[ 1-\sum_{i=1}^{N} u_{i}\left(1-\cos ^{i} \psi\right) \right ]
\end{equation}
where $\psi$ is the angle between emitted radiation and local radial vector (see Figure~\ref{fig:radiation_field}), and there is an implicit wavelength-dependence for $I(0)$ and the $u_i$ values.  As the intensity is cylindrically symmetric relative to the radial direction, the $J_{0}^{0}$ component (Equation~\ref{eqn:J00}) is
\begin{align}
    J_{0}^{0} &= \frac{1}{2} \int_{-1}^{1} I_{\mu} \mathrm{d}\mu 
\end{align}
where $\mu$ = $\cos\theta$.  $\psi$ is related to $\mu$ and the maximum value of $\theta$ in the integration as defined by when the rays are tangential to the stellar limb, denoted here as $\gamma$, which is used for consistency with the literature but should not be confused with $\gamma$ used prior to this point for the Euler rotation angle.  This relation is given by 
\begin{equation}
\cos \psi=\frac{\sqrt{\cos ^{2} \theta-\cos ^{2} \gamma 
}}{\sin \gamma 
}
\end{equation}
and, therefore,
\begin{align}
    J_{0}^{0} =\frac{I_{\mu=1}}{2} & \int_{\cos \gamma}^{1}\left[u_{0}(\nu) + \right. \nonumber \\
    & \left. \sum_{i=1}^{N} u_{i}(\nu) \frac{\sqrt{\left(\mu^{2}-\cos ^{2} \gamma\right)^{i}}}{\sin ^{i} \gamma}\right] \mathrm{d} \mu
\end{align}
This can be directly integrated, and for a limb-darkening law expanded to the quadratic terms, results in 
\begin{equation}
    J_{0}^{0} = \frac{I_{0}}{2} \left [ a_{0} + a_{1} u_{1} + a_{2} u_{2} \right ] \label{eqn:J00_sym_analytic}
\end{equation}
where $a_{n=0,1,2}$ are given by analytical functions of $\gamma$.  Using a similar approach, as also shown in  \citetalias{landidegl2004}, the analytical formula for $\omega$ (in the radial coordinate frame) can be written as:
\begin{equation}
w=\frac{1}{2} \frac{c_{0}+c_{1} u_{1}+c_{2} u_{2}}{a_{0}+a_{1} u_{1}+a_{2} u_{2}}.\label{eqn:omega_sym_analytic}
\end{equation}
We include the analytical formula for the $a_n$, $b_n$, $c_n$ values from \citetalias{landidegl2004} in Appendix~\ref{sec:appendix1}. 

\subsubsection{Integrated Signals in a Spherically Symmetric Model}

\begin{figure}
    \centering
    \includegraphics[width=0.95\columnwidth]{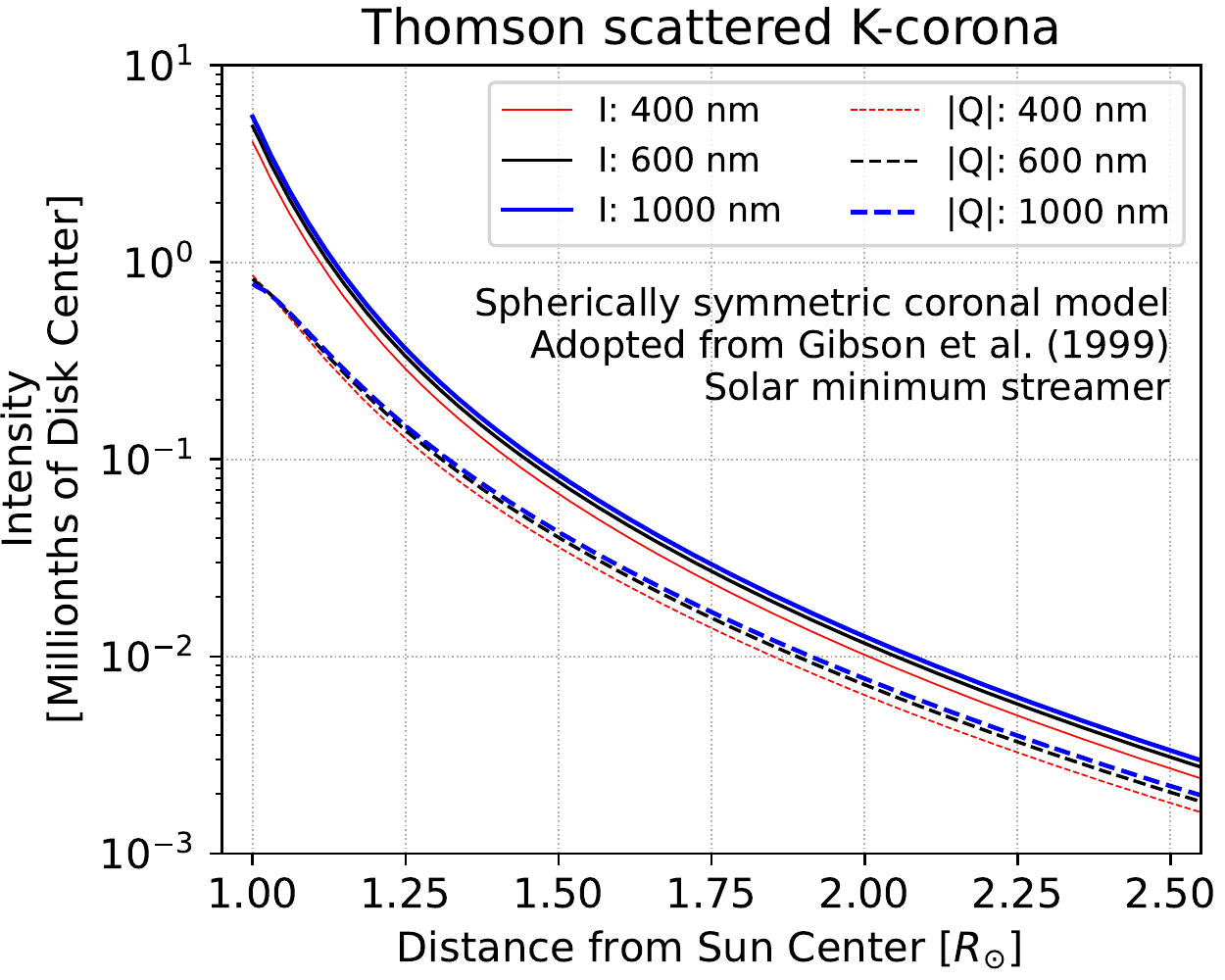} \\
    \includegraphics[width=0.95\columnwidth]{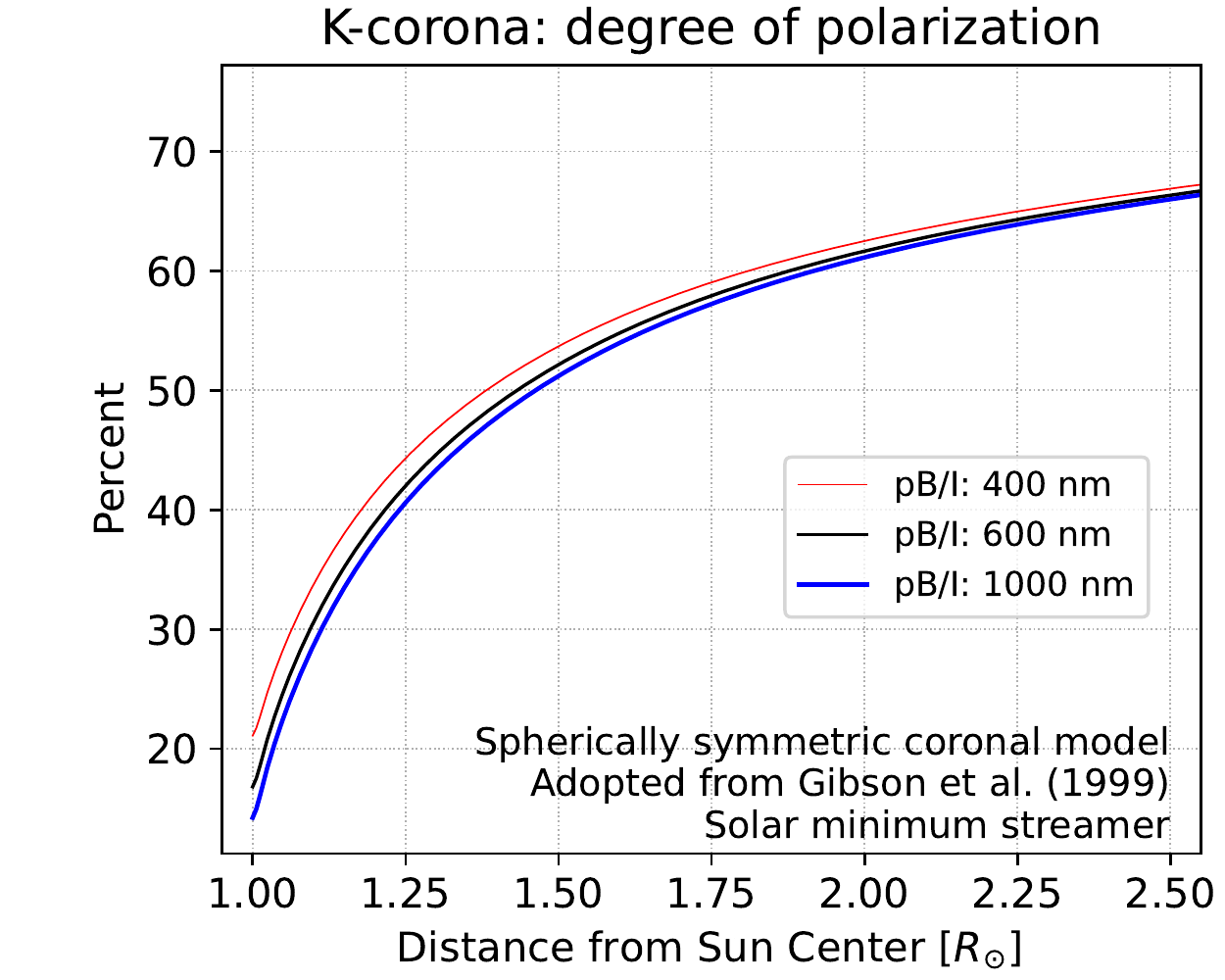}
    \caption{(top) Forward synthesized Thomson scattered K-corona intensity and the absolute value of Stokes Q for a spherically symmetric electron density model and cylindrically symmetric incident radiation field. (bottom) The corresponding degree of polarization.}
    \label{fig:model_sym}
\end{figure}

It can be easily demonstrated that Equations~\ref{eqn:epsI_tensor_sym} and~\ref{eqn:epsQU_tensor_sym}, together with Equations~\ref{eqn:J00_sym_analytic} and~\ref{eqn:omega_sym_analytic}, are consistent with the equations often used to compute the polarized Thomson scattered intensities in the solar case, as reviewed in \citet{howard2009}. One difference of the equations provided here is that the limb-darkening law is expanded to the quadratic terms, whereas \citet{howard2009} and earlier authors consider only the linear limb-darkening terms.  The quadratic expansion allows direct use of the wavelength-dependent quadratic expansion values for solar limb darkening as cataloged in Allen's Astrophysical Quantities \citep{cox2000}.  In general, we find the difference in the polarized Thomson scattered signal to be less than a few percent between the quadratic and best-fit linear expansion. 

In Figure~\ref{fig:model_sym}, we calculate, as verification of our method and for reference, the integrated Stokes signals for the Thomson-scattered K-corona for a spherically symmetric coronal model.  We use the empirical model of a solar streamer near solar minimum from \citet{gibson1999}.  The Stokes signals are calculated by integrating the emission coefficients along the line-sight, i.e.
\begin{equation}
S_{i} = \int_{s} \epsilon_{i} ds, 
\end{equation}
where $S_{i}$ is the Stokes vector $(i=\{I,Q,U,V\})$. The units are given in millionths of the disk center solar spectral radiance.  As the classical Thomson scattering cross section is not wavelength dependent, it is primarily the spectrally-dependent limb-darkening that results in the weak chromaticity of the resultant curves.  Note that we plot the absolute value of Stokes Q, as the sign of Q here is negative, referring to linear polarization oriented perpendicular to the projected radial vector.  The bottom panel shows the degree of polarization, given by the polarized brightness (pB) divided by the total intensity: 
\begin{equation}
    DoP = \frac{pB}{I} = \frac{\sqrt{Q^2 + U^2 + V^2}}{I}.
\end{equation}
As is evident, Figure~\ref{fig:model_sym} reproduces the K-corona characteristics described in the introduction.

\section[Forward Synthesis in the Non-Cylindrically Symmetric Case]{Forward Synthesis in the Non-Cylindrically \\ Symmetric Case}\label{sec:forward}

We now turn our focus to how active regions, through their influence on the local radiation field, alter the magnitude and orientation of the polarized Thomson scattered signal. \citet{Saint_Hilaire_2021} has recently performed a similar investigation wherein they use a numerical integration scheme to define `geometric factor maps' that account for the symmetry breaking characteristics in the radiation field above sunspots.  The benefit of the formalism outlined in the previous section is that such `geometric factor maps' become decomposed into an irreducible representation that can be rotated easily to different scattering directions.  Furthermore, we gain more intuition from using the (KQ) representation as the emissivities directly map to particular components of the expanded incident radiation field (see Equations~\ref{eqn:epsI_tensor1}-\ref{eqn:epsQU_tensor}).

\subsection[Radiation Field Calculations in a 3D MHD Active Region Simulation]{Radiation Field Calculations in a \\ 3D MHD Active Region Simulation} 

We forward synthesize the Thomson scattered spectral radiance emergent from an advanced 3D radiative magnetohydrodynamic simulation of a solar active region generated by the MURaM code \citep{rempel2017}.  The same simulation snapshot has been described in \citet{schad2020} and \citetalias{schad2021} and used to synthesize visible and infrared polarized emission lines in the non-relativistic quantum theory based on the atomic density matrix formalism.  The simulation domain extends over a volume of $98.304 \times 49.152 \times 49.152$ Mm $(1024 \times 512 \times 1024$ voxels) and includes a bipolar active region with two simulated sunspots connected by an arcade magnetic field simulated up to $\approx$41 Mm above the photosphere.  

\citetalias{schad2021} studied the symmetry-breaking in the incident radiation field near the simulated sunspots by forward synthesizing the 3D photospheric continuum radiation field.  By numerical integration of Equations~\ref{eqn:J00}-\ref{eqn:J22}, \citetalias{schad2021} determined the KQ components of the radiation field in the local solar radial frame.  Starting with these previously calculated $J_{Q}^{K}$ quantities, we here determine the relevant tensor components in the line-of-sight frame that contribute to the Thomson scattered signal.  These are $J_{0}^{\prime0}$ (=$J_{0}^{0}$), $J_{0}^{\prime 2}$, and $J_{2}^{\prime 2}$ as included in Equations~\ref{eqn:epsI_tensor1}-\ref{eqn:epsQU_tensor}.

\begin{figure*}
    \centering
    \includegraphics[width=0.95\linewidth]{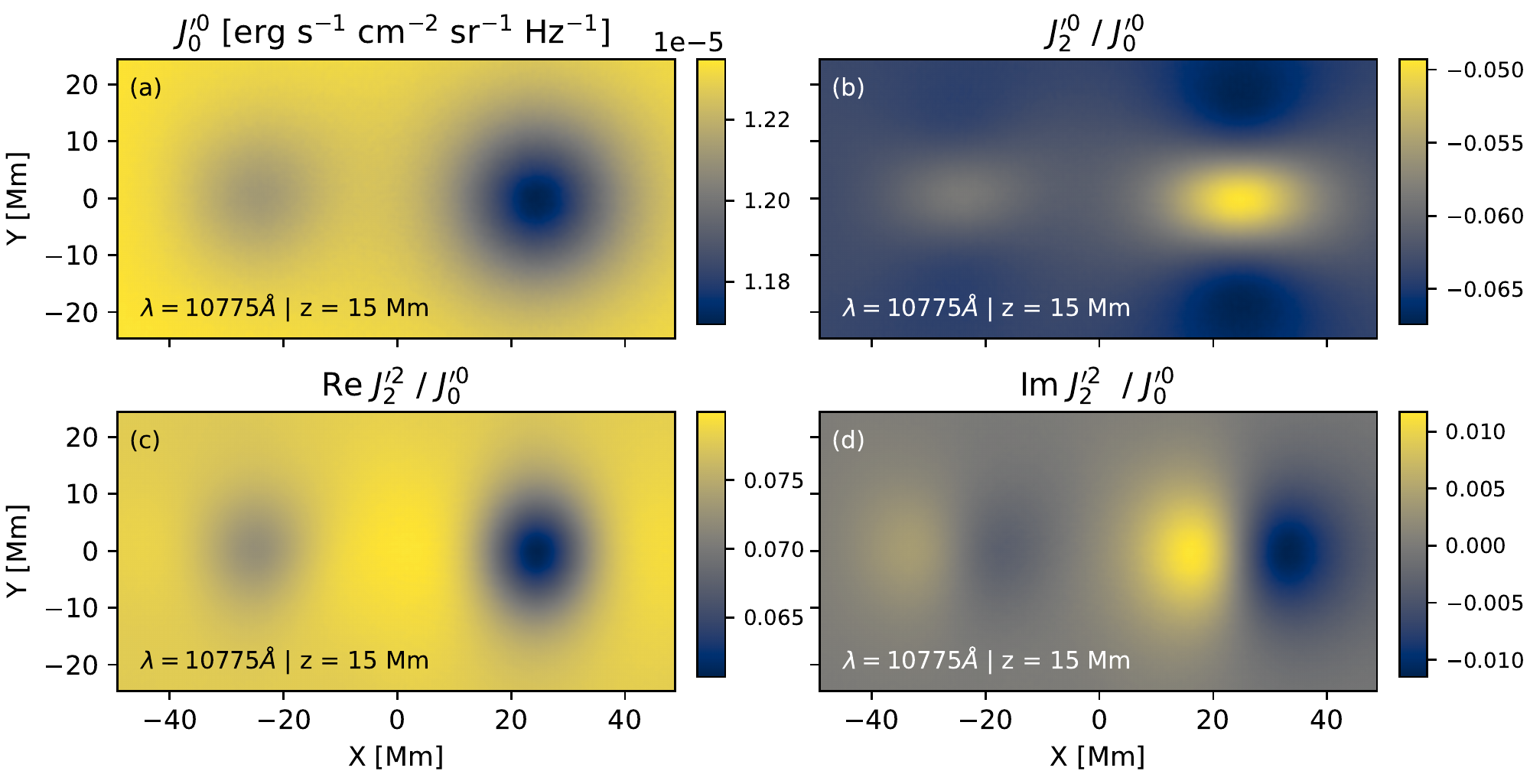} \\
    \includegraphics[width=0.95\linewidth]{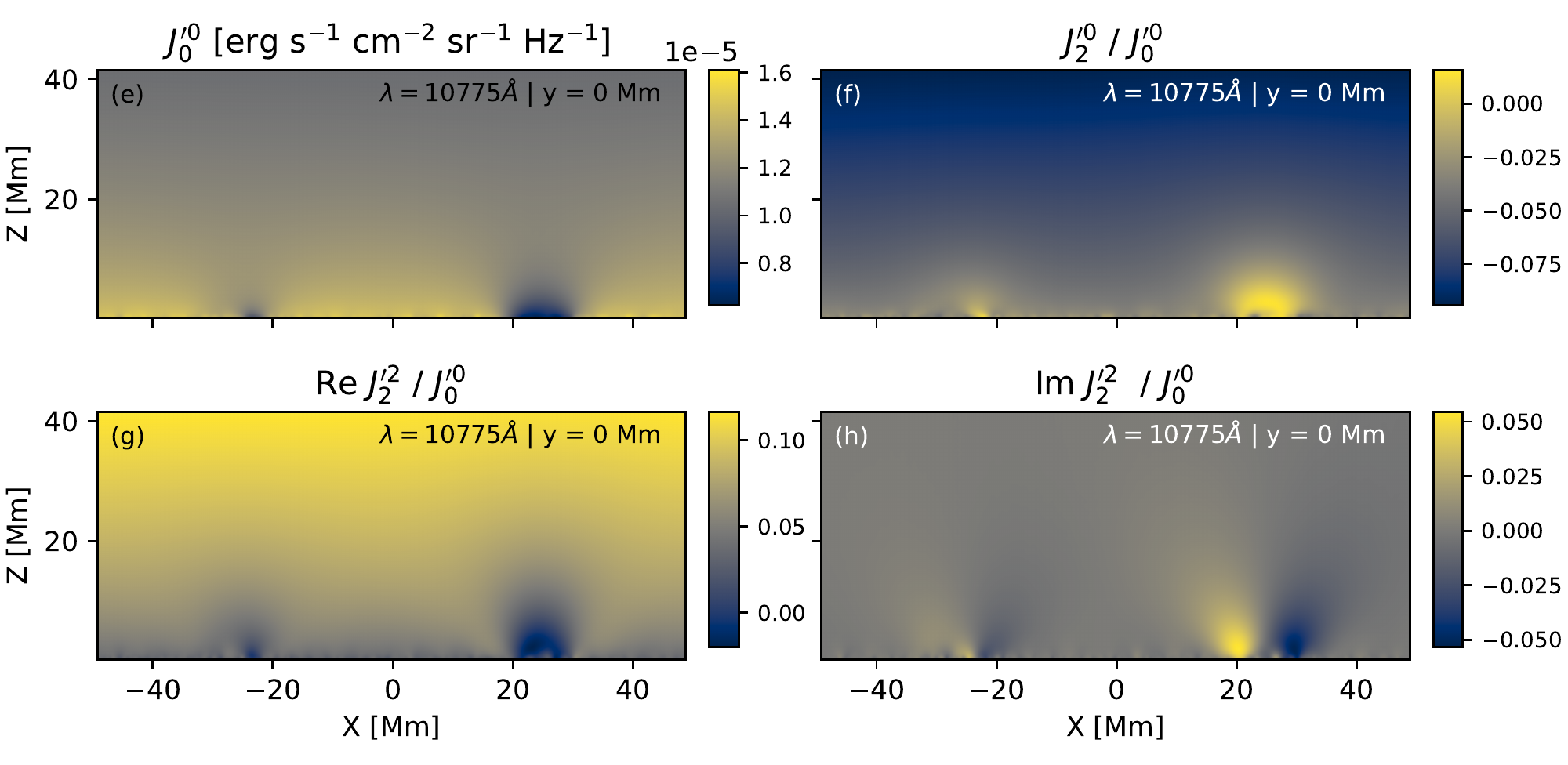} 
    \caption{The incident radiation field components within the bipolar 3D MURaM coronal simulation that contribute to the Thomson scattered continuum at $\lambda = 10775$ \mbox{\AA}.  The observer's line-of-sight aligns with the $Y$ axis, and for reference, two simulated sunspots are located near $X\pm25$ Mm ($Y=0$ Mm). The top panels (a-d) provide a horizontal slice at $Z=15$ Mm, and the bottom panels (e-h) show a vertical slice at $Y=0$ Mm.  Note the total mean intensity ($J_{0}^{\prime 0}$) is given in spectral radiance units while each other panel is normalized by $J_{0}^{\prime 0}$.}
    \label{fig:rad_field}
\end{figure*}

Figure~\ref{fig:rad_field} panels a-d and e-h, respectively, show the contributing components of the incident radiation field to the Thomson scattered continuum quantified in the scattering reference frame for a horizontal ($z=15$ Mm) and vertical ($y=0$ Mm) slice of the simulation.   As in \citetalias{schad2021}, the observer is assumed to be located in $-Y$ direction with a line-of-sight parallel to the Y axis, and the reference direction for Stokes +Q is aligned with the radial direction. The wavelength is set to be 10775 \mbox{\AA}, which is local to both the \ion{Fe}{13} 10746 \mbox{\AA} and 10798 \mbox{\AA} forbidden coronal emission lines. 

\begin{figure*}
    \centering
    \includegraphics[width=0.95\linewidth]{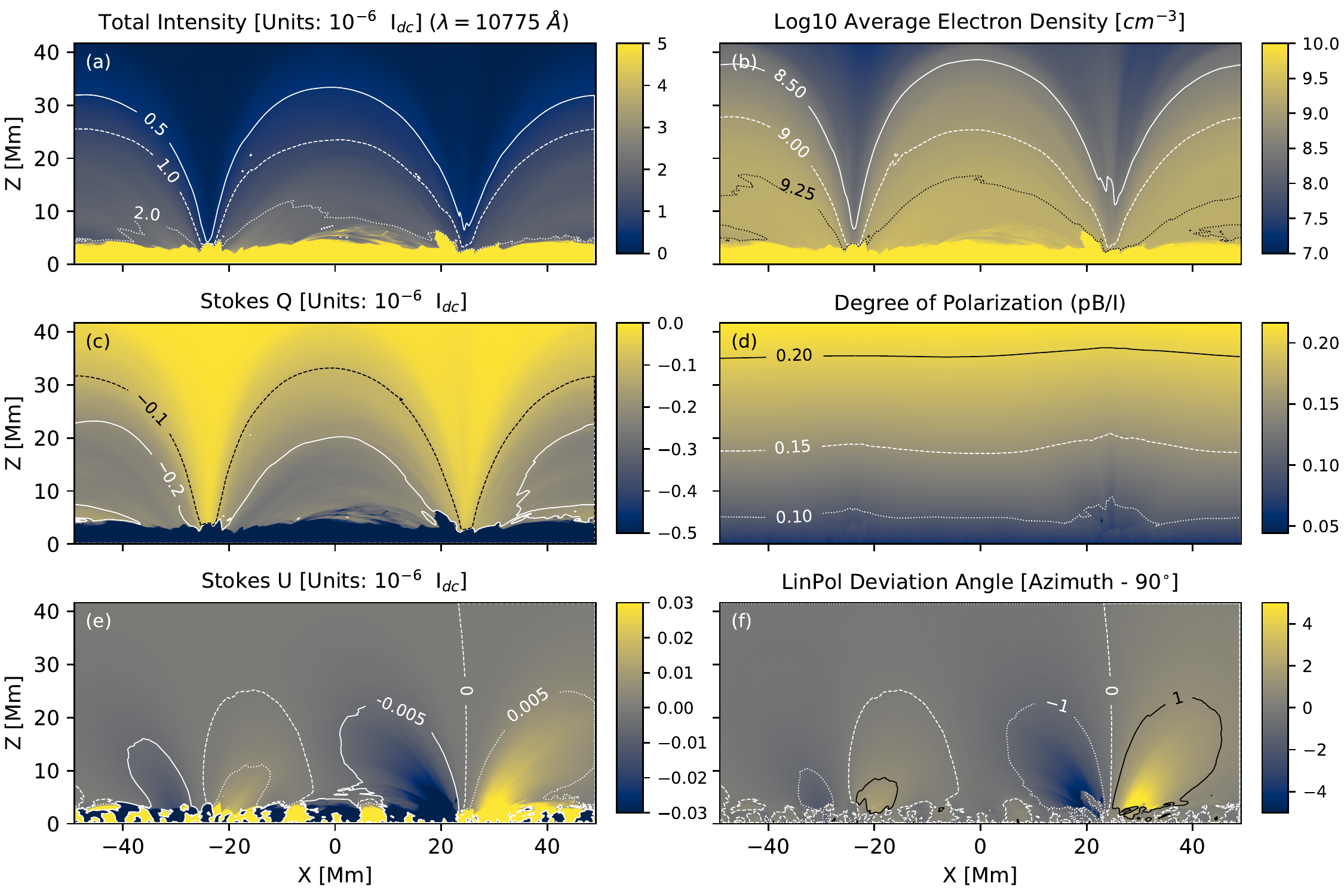}
    \caption{Synthesized polarized Thomson scattered continuum along the -y direction through the bipolar MHD coronal simulation.  Panels a,c,e show the total intensity, Stokes Q, and Stokes U intensities in units of millionths of the disk center spectral radiance.  Panel b is the line-of-sight average electron density in the simulation domain.  Panel d shows the degree of polarization while panel f shows the deviation angle of the linear polarization direction relative to the tangential direction, which in the establish reference frame is equal to the azimuth ($-0.5 \arctan2(U,Q)$) minus 90 degrees.}
    \label{fig:stokesThomson}
\end{figure*}

\subsection{Integrated Polarized Thomson Scattered Intensities}\label{sec:thomson_maps}

The line-of-sight integrated polarized Thomson signals at 10775 \mbox{\AA} are shown in Figure~\ref{fig:stokesThomson} and can be compared to the radiation field components in Figure~\ref{fig:rad_field}.  As expected, the Stokes Q signals are, with few exceptions, negative and larger than Stokes U.  Recall that Stokes U is zero for the cylindrically symmetric case, whereas here, the symmetric breaking effects of the two sunspots introduce an imaginary component in the quadrupole moment of the radiation field (see Fig.~\ref{fig:rad_field}d and h).  This results in the rotation of the plane of linear polarization relative to the expected tangential direction. As shown in Figure~\ref{fig:stokesThomson} panel f, the range of this deviation is approximately $\pm 5^{\circ}$ and is larger with closer proximity to the sunspots near $X = \pm25$ Mm. 
\begin{figure}
    \centering
    \includegraphics[width=0.95\linewidth]{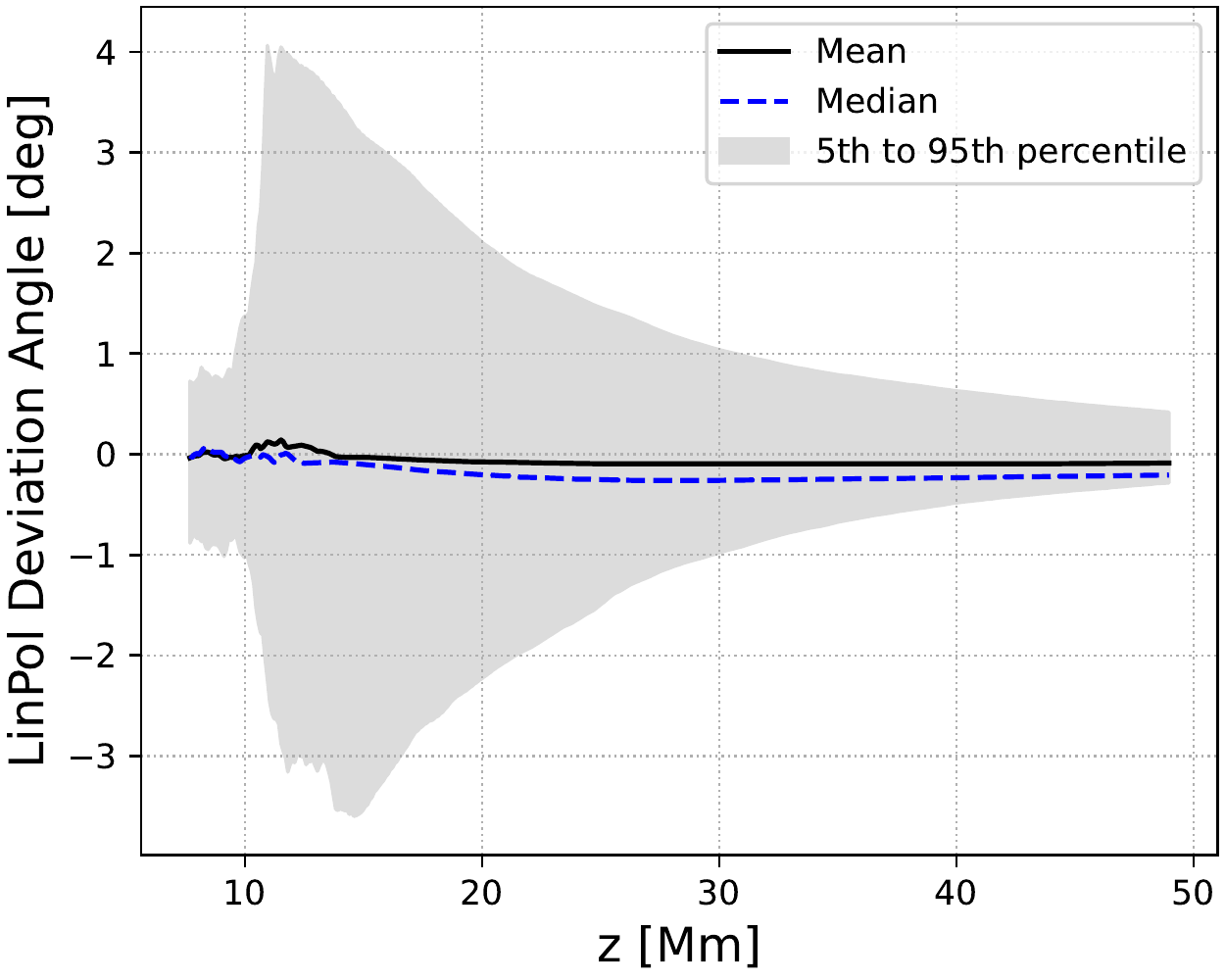}
    \caption{Height dependence of the linear polarization deviation angle relative to the tangential direction for the synthetic Thomson scattered signals shown in Figure~\ref{fig:stokesThomson}.}
    \label{fig:devAng}
\end{figure}

As discussed in the introduction, we make use (in the next section) of the expectation that the Thomson scattered K-corona polarization orientation is tangential to the solar limb.  Clearly, this is not fully satisfied by our synthetic observables due to the symmetry-breaking effects of the sunspots on the incident radiation field.  That said, as shown in Figure~\ref{fig:devAng}, the absolute mean and median deviation angle as a function of height is less than 0.15 and 0.25 degrees, respectively, while the 5th to 95th percentile values at the top of the domain are -0.42 to 0.29 degrees.  Thus, without more extensive modeling, one may significantly mitigate the role of the symmetry breaking on the expected continuum polarization direction by averaging over structures and/or avoiding the lower coronal regions above sunspots. 

\begin{figure}
    \centering
    \includegraphics[width=0.95\linewidth]{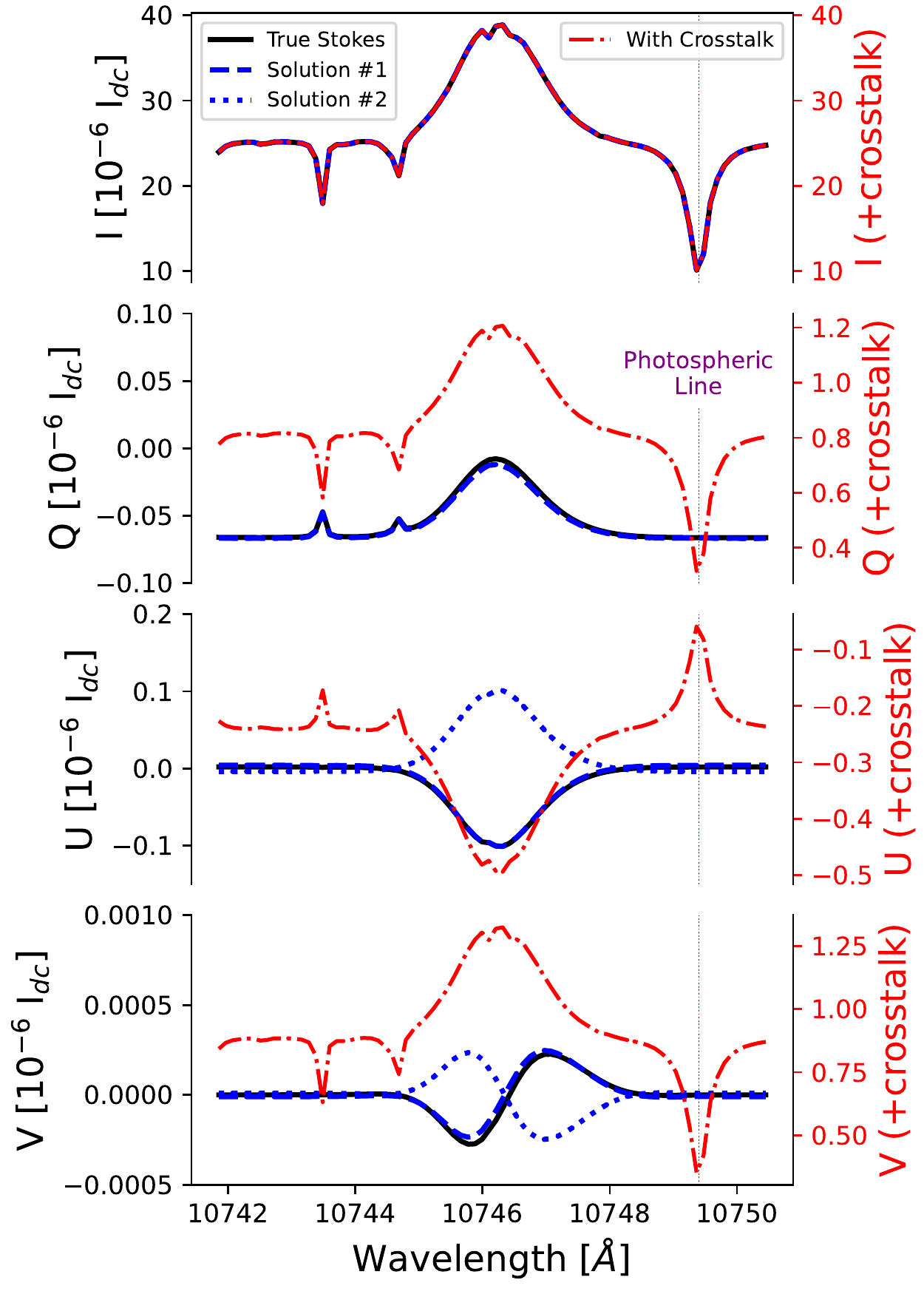} \\
    \includegraphics[width=0.95\linewidth]{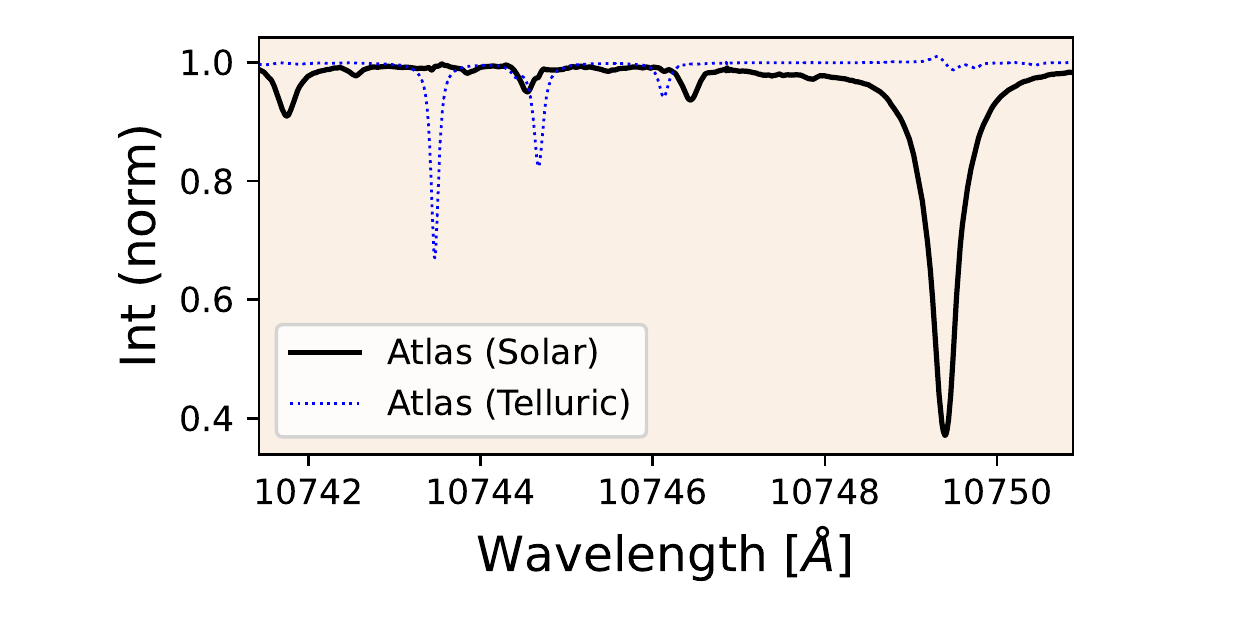} 
    \caption{(top four panels) The true synthetic Stokes vector emergent from the MHD simulation at $\langle x,z\rangle = \langle 32,32 \rangle$ Mm is given in black.  Dashdot red lines represent the `measured' Stokes vector with polarized crosstalk added, while the blue lines show the two solutions found for the crosstalk corrected signals.  The right axis tick marks apply only to the crosstalk added signals.  (bottom panel) The photospheric atlas spectrum acquired at disk center separated into solar and telluric components.}
    \label{fig:profiles}
\end{figure}

\begin{figure*}
    \centering
    \includegraphics[width=0.95\linewidth]{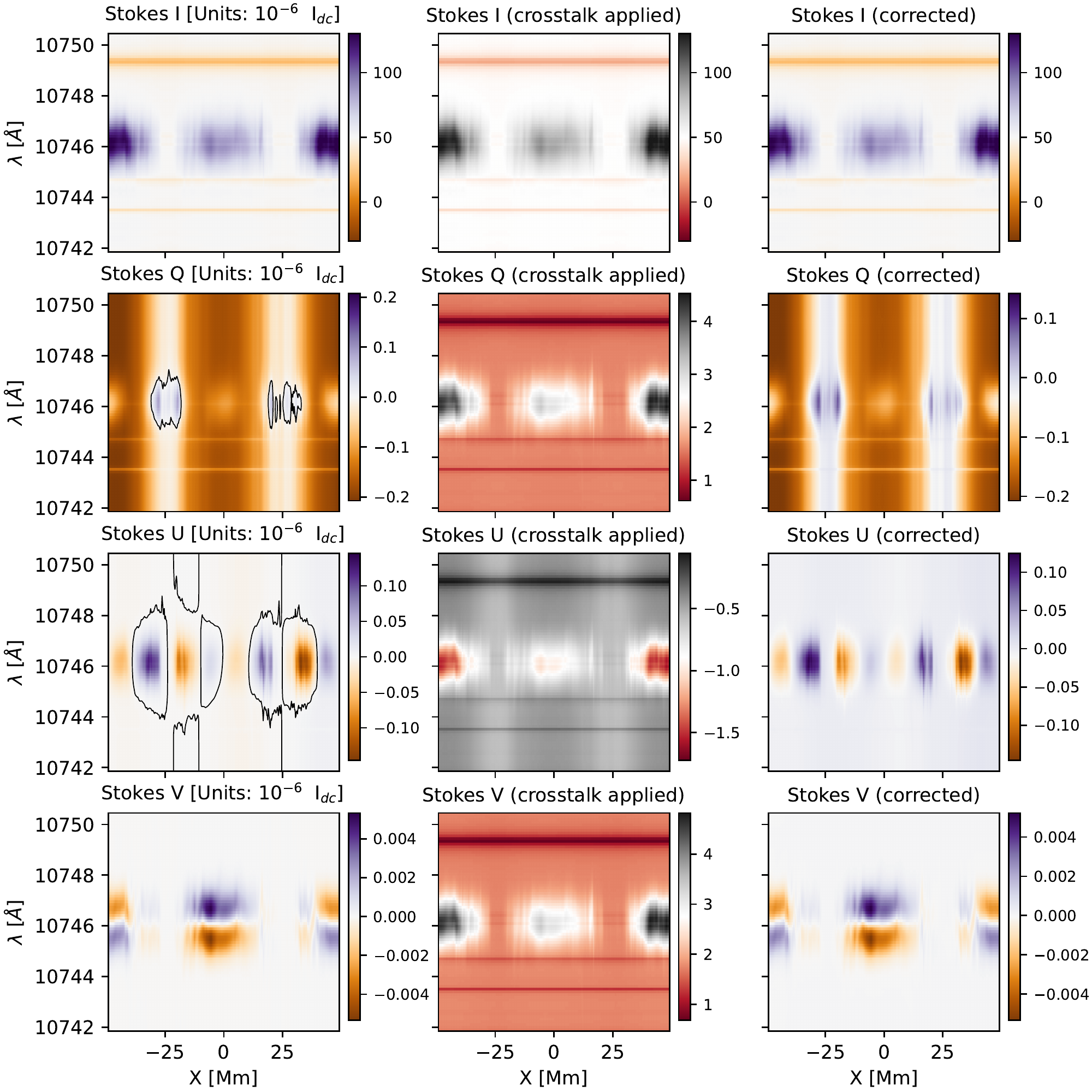}
    \caption{Synthetic Stokes spectra centered on the \ion{Fe}{13} 10746 \AA\ line including Thomson scattering and an unpolarized background spectrum.  The Stokes spectra incident on the optical system at a simulation height of Z = 30 Mm are shown on the left. Zero-level contours for Q and U are provided to distinguish regions where Stokes U takes negative or positive values. The middle column shows the Stokes spectra with crosstalk applied.   The right column shows spectra after applying the crosstalk correction method.}
    \label{fig:stokesSpec}
\end{figure*}

\section[An Ad Hoc Polarization Correction Technique for the Corona]{An Ad Hoc Polarization Correction \\ Technique for the Corona}\label{sec:adhoc}

Following the methodology introduced by \citet{jaeggli2022}, we now consider the use of the expected K-coronal and E-coronal polarized characteristics to constrain a model for an optical system's polarized response.  We begin with the assumption that the optical system is well described by a non-depolarizing Mueller matrix.  A Mueller matrix is a linear transformation matrix that maps one Stokes vector to another \citep[]{chipman2018}, and therefore we implicitly assume a set of polarimetrically modulated intensities has been demodulated in a consistent manner.  A non-depolarizing Mueller matrix can be further decomposed into an elliptical diattenuator and elliptical retarder, \textit{i.e.},
\begin{equation}\label{eqn:msys_decomp}
    \bm{M}_{sys} = \bm{M}_{P} \bm{M}_{R} = \bm{M}_{R} \bm{M}_{D}.  
\end{equation}
$\bm{M}_P$ and $\bm{M}_D$ are left-equivalent and right-equivalent diattenuators, respectively, and $\bm{M}_R$ is the Mueller matrix for a general elliptical retarder.  Below, we use the left-equivalent diattenuator form of this decomposition and further use the z-x-z extrinsic Euler angles to express the general elliptical retarder model, \textit{i.e.} 
\begin{align}
\bm{M}_{R}&(\alpha,\beta,\gamma) =  \nonumber \\
& 
\left[\begin{array}{cccc}
1  & 0  & 0  & 0  \\
0  & C_{\alpha} C_{\gamma}-C_{\beta} S_{\alpha} S_{\gamma} & -C_{\alpha} S_{\gamma}-C_{\beta} C_{\gamma} S_{\alpha} & S_{\alpha} S_{\beta} \\
0  & C_{\gamma} S_{\alpha}+C_{\alpha} C_{\beta} S_{\gamma} & C_{\alpha} C_{\beta} C_{\gamma}-S_{\alpha} S_{\gamma} & -C_{\alpha} S_{\beta} \\
0  & S_{\beta} S_{\gamma} & C_{\gamma} S_{\beta} & C_{\beta}
\end{array}\right]  \label{eqn:euler_zxz}
\end{align}
where $S_{\alpha}$ ($C_{\alpha}$) refers to the sine (cosine) of $\alpha$.  Once again, these Euler angles should not be confused with those defined in the previous section.  In this case, they can be interpreted as retardances.  

\subsection{Synthetic Coronal Spectra}

Let us now consider the components of a set of observable off-limb coronal Stokes spectra with the intent of generating model profiles with synthetic crosstalk applied. Similar to \citet{dima2019}, we write the true Stokes spectra incident on the optical system (subscript `i') as 
\begin{align}
I_{i} &= t_{atm} [I_{E} + I_{K} + I_{B}] \label{eqn:itot} \\
Q_{i} &= t_{atm} [Q_{E} + Q_{K}]  \\
U_{i} &= t_{atm} [U_{E} + U_{K}]  \\
V_{i} &= t_{atm} [V_{E}] \label{eqn:vtot}
\end{align}
where `E' and `K' represent the line-emissive and K-coronal components and $t_{atm}$ represents telluric atmospheric absorption.  All quantities have an implicit wavelength dependence. $I_{B}$ represents a `background' (B) component that is assumed to be unpolarized and which plays an important role in the ad hoc correction method described below.  $I_{B}$ consists not only of the F-corona, which we presume to be unpolarized, but also scattered light that is inherent to most coronagraphic observations of the low corona and typically dominates the F-corona signal. As discussed by \citet{dima2019}, both circumsolar forward scattering of solar disk light by aerosols in Earth's atmosphere and instrumental scattered light (due to diffraction and/or surface scattering by dust) can contribute to $I_{B}$, and they are generally indistinguishable from each other (and from the F-corona) without further modeling and/or external measurements.  As the telluric scattering occurs at shallow angles, it is assumed that the atmospheric scattered light remains unpolarized on-average, as it is comprised of integrated solar disk light.  We further assume the instrument induced scattered light follows (to good approximation) the same optical path as the coronal emission, which is expected to be a reasonable assumption for most coronagraphs where scattering in the entrance pupil is the dominate source.  In combination, these assumptions allow us to consider $I_{B}$ as a separate unpolarized source component with the same spectral characteristics as the integrated solar photospheric spectrum. 
In Figure~\ref{fig:profiles} (top four panels), we show one synthetic Stokes vector (black lines) emergent from the 3D MHD simulation at $\langle X,Z \rangle = \langle 32,32 \rangle$  Mm (see coordinates in Figure~\ref{fig:stokesThomson}) that combines the Thomson scattered signal with the \ion{Fe}{13} 10746 \mbox{\AA} polarized emission (computed in \citetalias{schad2021}) and a background/scattered light component as per Equations~\ref{eqn:itot}-\ref{eqn:vtot}.  The units are given in spectral radiance relative to the disk center intensity ($I_{dc}$).  $I_{B}$ is given a representative magnitude of 25 millionths of $I_{dc}$ with spectral features taken from a photospheric spectral atlas \citep{wallace1993}, which is shown separated into solar and telluric components in the fifth (or bottom) panel of Figure~\ref{fig:profiles}.  The most prominent spectral features include the \ion{Fe}{13} emission line near 10746 \mbox{\AA}, telluric H$_{2}$O lines at 10743.46 \mbox{\AA} and 10744.67 \mbox{\AA}, and the \ion{Si}{1} photospheric line at 10749.4 \mbox{\AA}.  The Thomson scattered emission vector at this location is $[I,Q,U,V]_{K} = [0.4011,-0.0662,0.0019,0] * 10^{-6} I_{dc}$; therefore, the linear polarization angle of the continuum is primarily tangential to the limb though rotated by $0.83^\circ$.  The \ion{Fe}{13} Q and U profiles have a Gaussian-like shape while Stokes V is anti-symmetric.  Note the presence of the telluric lines in the true Stokes Q signal and the absence of the scattered \ion{Si}{1} photospheric line signal at 10744.67 \mbox{\AA}. 
\subsection{Applying Polarimetric Crosstalk}

For illustrative purposes, we define one system polarized response matrix ($\bm{M}_{sys}$) for use in the rest of the paper.  We consider a weakly diattenuating system with a diattenuation magnitude D = 0.05 oriented at a polar angle of 45$^\circ$ and $-20^\circ$ azimuth on the Poincare sphere. The individual vector components are $\left \langle d_{H},d_{45},d_{R} \right \rangle = \left \langle 0.0332,-0.0121,0.0354 \right \rangle$.  The elliptical retardances, \textit{i.e.}, rotation angles, are set to $\left \langle \alpha, \beta, \gamma  \right \rangle = \left \langle 10^\circ, 30^\circ,-80^\circ \right \rangle$, which results in
\begin{align}
    \bm{M}_{sys}&|_{True} =  \nonumber \\
    & 
    \begin{bmatrix}
     \phantom{-}1.    &   \phantom{-}0.025 & \phantom{-}0.041 &  \phantom{-}0.015 \\
     \phantom{-}0.033 &   \phantom{-}0.319 & \phantom{-}0.809 &           -0.492 \\
               -0.012 &             -0.943 & \phantom{-}0.319 &            -0.087 \\
     \phantom{-}0.035 &   \phantom{-}0.087 & \phantom{-}0.493 &   \phantom{-}0.865
    \end{bmatrix}. \label{eqn:msys_true}
\end{align}
After applying this defined system Mueller system to the incident Stokes vector through matrix multiplication, the exit (or measured) Stokes vectors appear as shown by the dash-dotted red lines in Figure~\ref{fig:profiles}. With the addition of polarized crosstalk, the continuum polarization level is non-zero for each Stokes parameter and the scattered  \ion{Si}{1} line is now present in each as well.  Furthermore, the Gaussian-like coronal emission profile in the incident intensity dominates the spectral structure of all components such that there is no longer a recognizable anti-symmetric Stokes V signal.

\subsection{The Ad Hoc Correction Merit Function and its Application to a Single Profile}

Our goal is to use the \textit{a priori} constraints for the spectral content of the incident Stokes vectors to determine and remove the effects of the instrument's polarized response.  In contrast to \citet{jaeggli2022}, we here treat a case where the incident continuum is polarized, which entangles the diattentuation and retardance effects contained in Equation~\ref{eqn:msys_decomp} for the continuum.  Furthermore, as the coronal Q and U emission generally have a line shape that is similar to the total intensity, we cannot differentiate crosstalk between I and $(Q,U,V)$ from that between the polarized states themselves.  Instead, we must take advantage of the scattered photospheric line signal using a technique first introduced by \citet{dima2019}.  The presence of a solar absorption line in the measured (subscript `m') polarized spectra, when the polarization of the F-corona is negligible, directly indicates crosstalk from I to the polarized state.  Using Equation~\ref{eqn:msys_decomp}, we difference the measured polarization in the continuum ($\lambda = \lambda_{c}$) from that in the photospheric line ($\lambda = \lambda_{L}$) and obtain an inverse formulation for the diattenuation vector quantities in terms of the measured spectra: 
\begin{align}
d_{H} &= \frac{Q_{m}(\lambda_{L}) - Q_{m}(\lambda_{c})}{ I_{m}(\lambda_{L}) - I_{m}(\lambda_{c})}, \\
d_{45} &= \frac{U_{m}(\lambda_{L}) - U_{m}(\lambda_{c})}{ I_{m}(\lambda_{L}) - I_{m}(\lambda_{c})}, \\
d_{R} &= \frac{V_{m}(\lambda_{L}) - V_{m}(\lambda_{c})}{ I_{m}(\lambda_{L}) - I_{m}(\lambda_{c})}.
\end{align}
Following \citet{dima2019}, each can also be found by minimizing the difference for multiple points within a solar line, assuming an average value obtained in the continuum, \textit{e.g.}, 
\begin{align}
\underset{d_{H}}{\text{minimize}} \quad  \sum_{\lambda =\lambda_{L}}^{N_{L}} & \{d_{H} \left [ (I_{m}(\lambda) - I_{m}(\lambda_{c}) \right] \nonumber \\ 
& - [Q_{m}(\lambda) - Q_{m}(\lambda_{c})]\}^{2} 
\end{align}
Applying this directly to the `measured' profiles, \textit{i.e.}, the synthetic profiles with crosstalk added (Figure~\ref{fig:profiles}), using the \ion{Si}{1} line with $10749.58 < \lambda_{L} < 10750.45$ \AA\ (to avoid blending with the coronal line), we reproduce the input values for the diattenuation vector to high precision.  We can subsequently calculate the left-equivalent diattentuation matrix $\bm{M}_{P}$ and apply its inverse to the `measured' profiles, thereby removing the intensity to polarization crosstalk.

The crosstalk between the polarized states due to the system's elliptical retardance remains to be corrected at this stage.  Let us call this intermediary, diattenuation-corrected, Stokes vector $S_{R}$.  We now apply the assumption that the Thomson scattered continuum linear polarization is directed tangential to the projected stellar radius vector, which in the chosen reference frame implies Q is negative and U,V are both zero.  As an elliptical retarder (${\bf{M}_{R}}$) is non-depolarizing, we can directly measure the magnitude of continuum linear polarization using the $S_{R}$ vector, and associate it with the recovered (corrected) incident Q state:
\begin{equation}
    Q_{corr}(\lambda_{c}) = - \sqrt{Q_{R}(\lambda_{c})^{2} + U_{R}(\lambda_{c})^{2} + V_{R}(\lambda_{c})^{2}}
\end{equation}
where the negative sign accounts for the defined direction for Stokes +Q. 

In principle, using the determined value of $Q_{corr}(\lambda_{c})$ and the expressions for the 2nd column elements of ${\bf{M}_{R}}$ in Equation~\ref{eqn:euler_zxz}, one has enough information to infer the retardance angles $(\alpha,\beta,\gamma)$ using the measured continuum values of $S_{R}$.  However, the matrix elements are highly non-linear functions of these angles.  Small deviations in the true incident U continuum polarization and/or measurement noise makes this solution unstable. Therefore, we add another constraint that requires the line-integrated Stokes V signal to be zero, as discussed in the introduction. To summarize, we solve the following minimization problem: 
\begin{equation}
\begin{aligned}
 & \underset{\alpha,\beta,\gamma}{\text{minimize}}
 & & \left | Q_{corr}(\lambda_{c}) +  \right. \\
 &&& \left. \sqrt{Q_{R}(\lambda_{c})^2 + U_{R}(\lambda_{c})^2 + V_{R}(\lambda_{c})^2} \right | , \\
 &&& \left | U_{corr}(\lambda_{c})\right |,\\
 &&& \left | V_{corr}(\lambda_{c})\right |,\\ 
 &&& \left | \sum_{\lambda}V_{corr}(\lambda) \right | \\
 & \text{subject to}
 & & S_{corr} = \bm{M}_{R,FIT}^{-1} S_{R}.
\end{aligned}
\end{equation}
where $\bm{M}_{R,FIT}^{-1}$ is the inverse of the fitted elliptical retarder's Mueller matrix with free variables $(\alpha,\beta,\gamma)$. 

Using the Nelder-Mead minimization algorithm\footnote{We used the implementation of the Nelder-Mead algorithm provided in Scipy optimize.minimize module \citep{Gao2012,scipy2020}.} and many random initial guesses, we find the above technique yields two possible solutions,
\begin{align}
    \langle \alpha,\beta,\gamma\rangle_{1} & = \langle 8.42^\circ,   30.04^\circ, -80.0^\circ \rangle, \\
    \langle \alpha,\beta,\gamma\rangle_{2} & = \langle  171.58^\circ, 149.96^\circ, 100.0^\circ \rangle, 
\end{align}
both of which are shown in Figure~\ref{fig:profiles} using blue lines.  We can further assess the error of these two solutions by calculating the error matrix $\bm{E}$ as follows
\begin{align}
     \bm{E}_{FIT} &= \bm{M}_{sys}|_{FIT}^{-1} \bm{M}_{sys}|_{True} \nonumber \\
     & = \begin{bmatrix}
  1 & 0 & 0 & 0 \\
  0 & \pp0.9996 & -0.0276 & -0.0001 \\
  0 & \pm0.0304 & \pm0.9996 & \pm0.0007 \\
  0 & \mp0.0001 & \mp0.0007 & \pm1.0000 \\
\end{bmatrix} \label{eqn:err_single}
\end{align}
where the top (bottom) value of the $\pm$ and $\mp$ signals refers to solution \#1 (\#2).   The error magnitude is the same in both cases; only the signs of the 3rd and 4th row flip, which correspond to the corrected U and V profiles.  The sign degeneracy for U and V is expected as there is no \textit{a priori} constraint applied to their sign.  Instead, this degeneracy must be resolved with other methods.  The deviation of the error matrix from unity can be explained by the limits of our approximation.  Recall the plane of linear polarization deviates from our assumed direction, in this example, by $0.83^\circ$.  The equivalent Mueller rotation matrix for this deviation is given by
\begin{equation}
\begin{bmatrix}
  1 & 0 & 0 & 0 \\
  0 & \pp0.9996 & -0.0289 & 0 \\
  0 & 0.0289 & 0.9996 & 0 \\
  0 & 0 & 0 & 1 \\
\end{bmatrix},
\end{equation}
which closely resembles the error matrix above.  As such, this deviation angle of the true incident vector is the dominant error source in the method (ignoring noise and/or other systematics).

\subsubsection{Application Along a Horizontal Slice} 

As the error magnitudes discussed above scales primarily with the deviation of the linear polarization angle from the tangential direction, we can attempt to improve the performance of this method by an appropriate selection of measured vectors.  One strategy is to avoid areas above active regions where symmetry-breaking in the incident radiation field can be neglected.  Another strategy is to include a diversity of measured vectors across an active region within the minimization steps described above.  This requires that one can assume the optical system's response is uniform (within some accuracy specification) across the considered field-of-view, and that the field-of-view includes structure distributed about a symmetry-breaking feature such that the mean deviation angle is reduced, as in Figure~\ref{fig:devAng}.

The left column panels of Figure~\ref{fig:stokesSpec} show synthetic Stokes spectra that include the E, K, and background components for a horizontal slice across the MHD simulation at Z = 30 Mm.  The `measured' spectra (with crosstalk) resulting from the application of Equation~\ref{eqn:msys_true} are shown in the middle column. Using all spectral profiles in this slice for the ad hoc correction method, we find the following two solutions for the elliptical retardance angles:
\begin{align}
    \langle \alpha,\beta,\gamma\rangle_{1} & = \langle 10.6^\circ,   30.0^\circ, -80.^\circ \rangle, \\
    \langle \alpha,\beta,\gamma\rangle_{2} & = \langle  169.4^\circ, 150.0^\circ, 100.^\circ \rangle, 
\end{align}
with an error matrix given by
\begin{align}
     \bm{E}_{FIT} &= \bm{M}_{sys}|_{FIT}^{-1} \bm{M}_{sys}|_{True} \nonumber \\
     & = 
\begin{bmatrix}
  1 & 0 & 0 & 0 \\
  0 & 0.9999 & 0.0108 & 0 \\
  0 & \mp0.0108 & \pm0.9999 & -0.0002 \\
  0 & 0 & 0.0002 & \pm1 \\
\end{bmatrix}
\end{align}
In comparison to Equation~\ref{eqn:err_single}, we find the rotational error terms that interchange Q and U to be reduced by a factor of three, and all terms that interchange with V are reduced to $\le 2 \times 10^{-4}$. The corrected Stokes spectra are shown in the right column of Figure~\ref{fig:stokesSpec}.

\section{Summary and Discussion}

There are two primary features of the work presented here.  We first have provided the formulation for the polarized emissivities of Thomson scattering where the unpolarized incident radiation field is treated using irreducible spherical tensors.  As the established theory used for the synthesis of the polarized E-corona uses the (KQ) representation, we find it advantageous to utilize a comparable formulation for the K-corona, especially when working with large MHD simulations.  In this way, the components of the incident radiation field need only to be calculated in one reference frame, and can subsequently be transformed to account for multiple lines-of-sight.  We find this is an advantage over the approach of \citet{Saint_Hilaire_2021} and can likewise be used to quantify the radiation field above real observed features, as demonstrated by \citet{schad2015} and \citet{schad2021FIRS}.

Secondly, we have proposed an ad hoc technique for deriving the Mueller matrix of a non-depolarizing optical system based on the combined \textit{a priori} characteristics of the polarized E- and K- corona signals in the low corona (assuming the F-corona is unpolarized).  The technique performs well in the ideal case, and is primarily limited by the median deviation of the linear polarization angle from the assumed orientation.  We have not considered the role of measurement noise here nor specific outlier cases (such as when the emission line's Stokes vector does not contain signal in all polarized states); however, we do expect each could influence the relative error of individual terms of the inferred system Mueller matrix.  We note that the signal-to-noise requirements for a Stokes V measurement in an emission line are generally two to three orders of magnitude more stringent than Q and U, as observed by \citet{lin2004}. 

A prerequisite for the application of the proposed technique is a spectrally-adjacent photospheric line that is scattered by the sky, or the optical system, into the target optical path.  Importantly, we assume this scattered light can be treated as an unpolarized component of the total signal incident on the optical system and influenced by the same polarization response matrix.  Many important polarized coronal emission lines are located near a photospheric line \citep{schad2020, ali2022}; however, the application of our technique may in some cases be complicated by line blends and/or weak signals.  In the event a suitable photospheric line is unavailable, one may attempt to correct specific components of the crosstalk with other assumptions, at the cost of increased error.  \citet{dima2019} address the error incurred in measurements of linear polarization when one assumes the continuum is unpolarized (i.e. no K-coronal signal).  And \citet{lin2004} treat I,Q,U crosstalk into V by assuming I, Q, and U are proportional to each other and symmetric about line-center.

Together with adequate system knowledge of a given optical system to understand potential limitations, we believe the proposed technique is generally robust and useful, particularly for initial system characterization and/or validation of a more complete system model. The strategies discussed here are adaptable to both slit and filter-based spectropolarimeters, including those currently in operation or under development for DKIST and COSMO. 
\\
\\
\\
\noindent The National Solar Observatory (NSO) is operated by the Association of Universities for Research in Astronomy, Inc. (AURA), under cooperative agreement with the National Science Foundation. This research has made use of NASA’s Astrophysics Data System.

\appendix

\section{Integration Coefficients} \label{sec:appendix1}

The integration coefficients referenced in Section~\ref{sec:cyn_sym} are reproduced from \citetalias{landidegl2004} Section 12.3 here for convenience: 
\begin{equation}
\begin{aligned}
&C_{\gamma} = \cos \gamma \\ 
&S_{\gamma} = \sin \gamma \\ 
&a_{0}=1-C_{\gamma} \\
&a_{1}=C_{\gamma}-\frac{1}{2}-\frac{1}{2} \frac{C_{\gamma}^{2}}{S_{\gamma}} \ln \left(\frac{1+S_{\gamma}}{C_{\gamma}}\right) \\
&a_{2}=\frac{\left(C_{\gamma}+2\right)\left(C_{\gamma}-1\right)}{3\left(C_{\gamma}+1\right)} \\
&b_{0}=\frac{1}{3}\left(1-C_{\gamma}^{3}\right) \\
&b_{1}=\frac{1}{24}\left(8 C_{\gamma}^{3}-3 C_{\gamma}^{2}-2\right)-\frac{1}{8} \frac{C_{\gamma}^{4}}{S_{\gamma}} \ln \left(\frac{1+S_{\gamma}}{C_{\gamma}}\right) \\
&b_{2}=\frac{\left(C_{\gamma}-1\right)\left(3 C_{\gamma}^{3}+6 C_{\gamma}^{2}+4 C_{\gamma}+2\right)}{15\left(C_{\gamma}+1\right)}, \\
&c_{0}=3 b_{0}-a_{0} \\
&c_{1}=3 b_{1}-a_{1} \\
&c_{2}=3 b_{2}-a_{2} \nonumber
\end{aligned}
\end{equation}

\bibliography{main}{}
\bibliographystyle{aasjournal}

\end{document}